\documentclass[sigplan]{acmart}
\usepackage{graphicx}

\usepackage{acronym}

\usepackage{mdframed}
\usepackage{framed}

\usepackage{balance}
\usepackage{hyperref}
\usepackage{mathptmx}
\usepackage{amsmath}
\usepackage{color,soul}

\AtBeginDocument{%
  \providecommand\BibTeX{{%
    \normalfont B\kern-0.5em{\scshape i\kern-0.25em b}\kern-0.8em\TeX}}}

\copyrightyear{2023} 
\acmYear{2023} 
\setcopyright{acmlicensed}\acmConference[NSPW '23]{New Security Paradigms Workshop}{September 18--21, 2023}{Segovia, Spain}
\acmBooktitle{New Security Paradigms Workshop (NSPW '22), September 18--21, 2023, Segovia, Spain}
\acmPrice{15.00}
\acmDOI{xxx}
\acmISBN{xxx}

\acrodef{ExD}{\emph{Explainable Deletion}}
\newcommand{\etal}[0]{et~al{.}}

\begin{document}

\title{ExD: Explainable Deletion}

\author{Kopo M. Ramokapane}
\email{marvin.ramokapane@bristol.ac.uk}
\orcid{0000-0001-8420-3929}
\authornotemark[1]
\affiliation{%
  \institution{University of Bristol}
  \streetaddress{75 Woodland Road}
  \city{Bristol}
  \country{United Kingdom}
  \postcode{BS8 1UB}
}

\author{Awais Rashid}
\email{awais.rashid@bristol.ac.uk}
\orcid{0000-0002-0109-1341}
\affiliation{%
  \institution{University of Bristol}
  \streetaddress{75 Woodland Road}
  \city{Bristol}
  \country{United Kingdom}}

\renewcommand{\shortauthors}{Ramokapane and Rashid}

\begin{abstract}

  This paper focuses on a critical yet often overlooked aspect of data in digital systems and services\textemdash \emph{deletion}. Through a review of existing literature we highlight the challenges that user face when attempting to delete data from systems and services, the lack of transparency in how such requests are handled or processed and the lack of clear assurance that the data has been deleted. We highlight that this not only impacts users' agency over their data but also poses issues with regards to compliance with fundamental legal rights such as the \emph{right to be forgotten}. We propose a new paradign -- \emph{explainable deletion} -- to improve users' agency and control over their data and enable systems to deliver effective assurance, transparency and compliance. We discuss the properties required of such explanations and their relevance and benefit for various individuals and groups involved or having an interest in data deletion processes and implications. We discuss various design implications pertaining to explainable deletion and present a research agenda for the community.
\end{abstract}




\maketitle


\section{Introduction}
\label{Sec:Intro}

Modern technologies and services rely heavily on data-intensive systems driven by algorithms operating on large datasets. The collection and utilization of digital data have become indispensable for numerous businesses and services, as many aspects of life now depend on it. However, the mishandling and improper use of data can pose various privacy risks. Consequently, there has been a significant focus on ensuring the proper and ethical use of data in various aspects of these ecosystems, such as AI and the need for explainable AI. While this is crucial, there are other areas where explainability holds importance, particularly concerning the exercise of fundamental rights in regulations like GDPR, such as the right to be forgotten. Currently, aspects related to data deletion are often poorly explained or even missing entirely~\cite{ramokapane2022users}. However, a lack of understanding about deletion processes or where to find deletion controls may lead users to abstain from using certain systems or engage in self-censorship out of fear of being targeted or marginalized.

There is a lack of explainability surrounding data deletion, despite its significant implications for users' privacy and choice. For instance, understanding the completeness of the deletion procedure can help users make informed decisions and provide consent when deciding whether to use a system. Moreover, knowing how data is deleted can also influence how they interact with the system~\cite{clark2015saw,ramokapane2022users}. The argument in favor of explainability in systems often arises from the recognition that current systems are opaque, making it challenging to comprehend the complete processes and make informed decisions. In the context of data deletion, an ``opaque system'' refers to a system where the mechanisms or procedures that translate user inputs (e.g., controls used for data deletion requests) into outputs (e.g., data removal) are not visible to users. Despite data deletion sharing a similar level of opacity with other systems, such as AI systems, explainability in the context of deletion remains largely unexplored. This paper explores the potential means of achieving explainability in the context of data deletion.

According to the Oxford Dictionary, an explanation is ``a thing which explains, makes clear, or accounts for something; a statement that makes things intelligible.'' Lombrozo argues that explanations serve as the medium through which we exchange beliefs~\cite{lombrozo2006structure}. While explainability in AI systems may have existed as long as AI systems~\cite{Newell1958}, it has recently gained popularity. Explanations are seen as a means to achieve transparency and accountability. The AI literature posits that explanations enable users (or humans) to understand and interpret the system, providing insights into the rationale behind certain decisions and predicting their consequences. This perspective adopts a goal-oriented approach, focusing on the intended achievements of explanations. Consequently, the main goal of an explanation is to attain interpretability and comprehension of a system and its decision-making processes. This suggests that explanations should include pertinent information that is easy to understand.

Considering these arguments, we propose a novel paradigm in security known as~\ac{ExD}. The core objective of explainable deletion is to enhance transparency and accountability around data deletion and facilitate user understanding of data deletion processes. This approach empowers users to make informed decisions regarding systems and how they handle data deletion. 

Current approaches to deletion often leave users with limited or no understanding of the actual consequences of their actions~\cite{Ramokapane2017Feel,Murillo2018Press}, the extent or completeness of the deletion process~\cite{reardon2013sok,Murillo2018Press,khan2018forgotten,ramokapane2022users}, and the potential risks of system failures in effectively removing their data~\cite{reardon2013sok,zhu2013velocity,schnitzler2021sok}. By addressing the need for explainability in data deletion, we aim to contribute to advancing security practices, and fostering user trust in technology systems. Prior works~\cite{Ramokapane2017Feel,Murillo2018Press,ramokapane2022users} show that aspects of deletion are often missing yet users desire to know some of them. Moreover, Liu et al.~\cite{liu2022your} and Habib et al.~\cite{habib2020s} showed that when users are given extra bit of information on deletion their privacy-related responses change. By providing explanations, users' mental models can be enhanced, the gap between the ``actual data deletion process'' and the ``perceived deletion process'' can be bridged. For service providers, the realization of explainable deletion holds significant potential in fostering trust between them and their users. Furthermore, it facilitates accountability in data deletion practices vis-à-vis regulatory requirements. By shedding light on the various aspects of data deletion, service providers can cultivate a sense of trust and reliability in their users, assuring them that their data will be appropriately managed and safeguarded.

While our work focuses on data deletion, we also acknowledge that the principle of explainability holds significant importance for other facets of the data cycle. The tenet can be applied to other data management aspects such as data storage, portability, archiving, de-identification, and accessibility. Data deletion should not be perceived as an isolated process but rather intricately connected to other facets of the data lifecycle. In fact, understanding the explainability of deletion can have profound implications for broader data management processes and may help contribute towards Equitable Privacy.

Our work focuses specifically on explainable deletion because we recognize a significant gap in the literature concerning this critical aspect of the data lifecycle. While there is no doubt that a holistic view of explainability is essential, it is also important to address individual components in-depth to develop a more comprehensive understanding of the data management process. Addressing the explainability of data deletion is a foundational step toward tackling more complex challenges in the data lifecycle.

This paper makes the following contributions:
\vspace{-4.5pt}
\begin{itemize}
    \item \textbf{Introduction of Explainable Deletion:} We introduce the concept of Explainable Deletion as a novel approach to enhance transparency in data deletion processes. We highlight the significance of transparency in addressing the challenges associated with data deletion.
    \item \textbf{Exploration of the Benefits of Explainable Deletion:} We shed light on the potential benefits of Explainable Deletion in addressing data deletion challenges. We emphasize the importance of deletion transparency in fostering trust, ensuring accountability, and empowering users to make informed decisions about their data.
    \item \textbf{Framework for Implementing Explainable Deletion:} We provide a framework for implementing Explainable Deletion. This framework offers design considerations for practitioners and developers to effectively incorporate transparency mechanisms into data deletion processes.
    \item \textbf{Provision of a Research Agenda:} We identify key research areas that require further investigation to advance the understanding and practical application of Explainable Deletion. This research agenda aims to promote deletion transparency in systems and guides researchers and practitioners to contribute to the ongoing development of Explainable Deletion.
\end{itemize}

The rest of the paper proceeds as follows. Section~\ref{Sec:Dimensions} discusses the dimensions of Explainable Deletion, utilizing examples to demonstrate specific situations that emphasize the importance and relevance of~\ac{ExD}. In Section~\ref{Sec:StateOfTheArt}, we review the current state of the art on data deletion, while Section~\ref{Sec:NewParadigm} introduces the concept of ~\ac{ExD}, including its definition, the aspects of Deletion, and the beneficiaries of~\ac{ExD}. Section~\ref{Sec:HowTo} delves into design considerations for Explanation Deletion, outlining what designers should prioritize. In Section~\ref{Sec:Operationalising}, we illustrate with an example how various aspects of deletion can help realize dimensions of~\ac{ExD}. Section~\ref{Sec:Agenda} presents a research agenda for the community, and finally, we conclude in Section~\ref{Sec:Conclusion}.

\section{Dimensions of Explainability}
\label{Sec:Dimensions}
In this section, we present illustrative examples that highlight various dimensions of explainable deletion. These examples are intended to demonstrate specific situations or scenarios emphasizing the importance and relevance of explainable deletion in practice. While multiple dimensions may be applicable in each case, we utilize each example to emphasize the significance of a particular dimension.

\subsection{Agency}

\begin{mdframed}[backgroundcolor=gray!10,linewidth=0.1pt,font=\small\sffamily]
\paragraph{Instant messaging apps}
In 2021, WhatsApp introduced a ``view once''~\footnote{https://faq.whatsapp.com/1077018839582332} feature for photos and videos. With this feature, the sender can mark a photo as view once, and the photo will disappear after the receiver has opened it. While this type of media sharing promises privacy, it remains unclear whether the disappearance means the deletion is permanent or not. Another feature offered by WhatsApp allows users to delete messages for themselves or everyone involved in the conversation. However, when users choose the ``Delete for me'' option, they lose control over the message they have sent. If they later desire to delete the message for everyone, they will no longer have access to it, though other participants in the chat still have access. The consequences of the deleting messages in this manner is not made clear or well-communicated to users~\cite{schnitzler2020exploring}.

\end{mdframed}
\vspace{5pt}
Providing explanations regarding data deletion serves multiple purposes. Firstly, it empowers users with a comprehensive understanding of the available mechanisms for deleting data and the processes involved during and after deletion. By knowing where these mechanisms are located, users can effectively exercise their right to delete data. Secondly, explanations will assist users in developing a deeper understanding of how different systems handle data deletion and the extent to which data is deleted. This is particularly significant as users often possess incomplete mental models of deletion. Moreover, by gaining insights into the variations among systems regarding deletion procedures and the scope of data deletion, users can make more informed choices regarding their privacy. This includes decisions concerning selecting or avoiding specific systems or services that align with their desired level of data deletion or privacy through deletion.

\subsection{Assurance}

\begin{mdframed}[backgroundcolor=gray!10,linewidth=0.1pt,font=\small\sffamily]
In 2017, there were reports that some deleted files and folders from Dropbox~\footnote{\textbf{Statement from Dropbox}: \textit{``A bug was preventing some files and folders from being fully deleted off of our servers, even after users had deleted them from their Dropbox accounts. While fixing the bug, we inadvertently restored the impacted files and folders to those users’ accounts. This was our mistake; it wasn’t due to a third party and you weren’t hacked."}} unexpectedly reappeared in users' accounts~\cite{dropbox2017}. According to the reports~\cite{dropbox2017,Catalin2017Dropbox}, some of these files had been deleted for over six years, despite Dropbox's stated retention policy of 30 days. The risks around the method used to delete data from Dropbox were not fully disclosed to users. The procedures handling data deletion are usually undisclosed to users, leaving them uncertain about the specific mechanisms employed and the outcomes to expect following the deletion of their data.
\end{mdframed}
\vspace{5pt}
The actual workings of data deletion usually take place away from users' eyes. In most cases, they only get to see the outcome of the deletion process. Consequently, they usually rely on trust that the system or service provider has fulfilled their commitment to delete their data. By explaining and being transparent about the procedures and outcomes associated with data deletion, users can obtain a more comprehensive understanding of the underlying mechanisms at work. This, as a result, will foster a sense of assurance and trust that their data is managed responsibly and ethically. Moreover, providing feedback may act as verification or reassurance, confirming that the users' intended actions have been fulfilled as expected. 

\subsection{Control}
\begin{mdframed}[backgroundcolor=gray!10,linewidth=0.1pt,font=\small\sffamily]
\paragraph{Accidental Deletion}
In 2014, a Dropbox user named Jan ˇCurn lost 8,343 photos and videos due to accidental deletion~\cite{Curn2014DropboxBug}. Jan discovered this unfortunate incident two months later while searching for a presentation for his Ph.D. thesis defense. Despite seeking help from Dropbox, he was informed that he had deleted his files several months ago, and since the retention period had expired (60 days). Unfortunately, most of the files were permanently deleted. The accidental deletion occurred when Jan attempted to desynchronize some folders to create space in his computer, but his computer Dropbox client crashed. Upon restart, the client considered all files deleted and removed them from the server.
\end{mdframed}
\vspace{5pt}
Explainable deletion affords users freedom and control over their deletion actions while interacting with systems. Users often inadvertently delete their data, and it is crucial for them to be aware of the available options for recovering from such mistakes. By providing explanations regarding possible courses of action, such as data recovery, their anxieties can be alleviated, and they can continue to explore and utilize the system with confidence. Moreover, understanding the various types of deletions available to them will give them the autonomy to choose the deletion type that is right and beneficial to them and their businesses. Understanding deletion can also act as a proactive strategy for users; it will help users reflect on their choices and behaviour before choosing or while interacting with various technologies. Overall, explaining deletion will help users understand how much control they have over their data while using the system.

\subsection{Compliance}
\begin{mdframed}[backgroundcolor=gray!10,linewidth=0.1pt,font=\small\sffamily]
Privacy regulations, for example, the General Data Protection Regulation (GDPR) and the California Consumer Privacy Act (CCPA) afford users (i.e., data subjects) the fundamental right to erasure. This right empowers users to request the deletion of their personal information from service providers. However, despite being able to exercise this right, users often encounter significant challenges in understanding the process of initiating such requests. Crucial guidance about the specific methods for requesting data erasure is frequently either absent or deliberately obscured, leaving users uninformed about the necessary steps to take.
\end{mdframed}
\vspace{5pt}

Regulations around deletion, for example, GDPR Article 17 requires service providers to provide processes and mechanisms to ensure that they can fulfil users' data deletion requests promptly and effectively. Therefore, explaining deletion can demonstrate that service providers are committed to complying with deletion regulations and have measures to handle deletion requests from their customers. Furthermore, explainable deletion will assist service providers in meeting users' right to be informed. Lastly, by being transparency about data deletion practices, service providers can demonstrate their commitment to regulatory compliance.

\subsection{Transparency}
\begin{mdframed}[backgroundcolor=gray!10,linewidth=0.1pt,font=\small\sffamily]
\paragraph{Smart home technologies}
Smart home technologies present a complex ecosystem to users, encompassing interconnected devices and involving multiple parties~\cite{clark2015saw,abdi2019more}. The usage of these devices entails the sharing or storing of data by various parties, often using various technologies, such as the cloud, to serve a range of purposes. Data deletion processes in this context are not always confined to the devices themselves but may instead happen through web or mobile app interfaces. However, in most cases, how data is deleted, the specific data being deleted, retention policies and the underlying mechanisms for deletion remain undisclosed and non-transparent to users.
\end{mdframed}
\vspace{5pt}
Deleting data from complex systems, such as smart home technologies and cloud computing systems, is a multifaceted task that often requires the execution of multiple processes. Ensuring complete data deletion from such ecosystems is inherently challenging. For instance, data stored in the cloud can be distributed across various layers (including logical layers) and multiple servers located worldwide. Consequently, when a deletion request is made, the deletion process must propagate through all these layers and services to remove the data effectively. Unfortunately, this process is generally unknown to users. Explaining (e.g., whitebox explanations~\cite{herlocker2000explaining}) such a process would not only assist users in comprehending that their data may not be entirely deleted but also help them understand the reasons behind this limitation. Furthermore, explaining the deletion process can help service providers show that their actions are not malicious or compromising users' data.

\section{Current State of the Art}
\label{Sec:StateOfTheArt}

\subsection{Nominal Data Deletion}
Modern computing systems allow users to delete their data when deemed unnecessary. The deletion process typically involves user-initiated actions, such as selecting the desired item for deletion and executing the corresponding command. In the case of computer systems, the deleted item is often moved to the trash can or recycle bin. If the user wishes to delete the item permanently, they can select it from the recycle bin and use the ``delete permanently'' option.

However, while the deleted item may no longer be visible, it is technically not fully deleted and can still be recovered~\cite{diesburg2010survey}. When a user requests the deletion of an item from the recycle bin, the item is first marked as deleted and made inaccessible through any user interface. The system then proceeds to remove the file from the storage medium. Depending on the medium used, the system may mark the memory blocks previously occupied by the data as available for reuse. The item's metadata is also updated to ensure it is no longer linked to other data. However, the entire contents of the deleted item remain in the system. The complexity of the process increases when data is replicated across multiple systems or accessible by multiple devices. Regardless of the system, remnants of the supposedly deleted data may be left behind, and using certain tools makes it possible to recover the supposedly ``deleted'' data~\cite{diesburg2010survey,reardon2013sok,shu2017data}.

\subsection{Assured deletion}
To address the limitations of nominal deletion and facilitate complete data deletion, researchers have proposed various methodologies. Some approaches~\cite{hughes2009disposal,reardon2013sok,reardon2014secure} focus on the physical destruction of the storage medium, which provides a high degree of assurance but is expensive and rarely employed. Alternatively, other techniques~\cite{diesburg2010survey,luo2016enabling} involve overwriting the memory blocks previously used for storing the data or filling the storage media with new insensitive data, thus concealing the original data. Other strategies~\cite{tang2010fade,rahumed2011secure,tang2012secure} concentrate on rendering the data unusable or inaccessible. Rather than removing the data from the media, it is encrypted, and the encryption key is securely disposed of. Another way to satisfy deletion requirements without removing data is through privacy concepts such as differential privacy, where identifiable data is removed or disassociated from the dataset~\cite{ginart2019making,sekhari2021remember}. In the case of protecting ``deleted'' tweets, Minaei~\etal{}~\cite{minaei2021deceptive} developed a tool to hide damaging tweet deletions. Paul and Saxena~\cite{paul2010proof} proposed a scheme to proof deletion in the cloud. Regarding emails, Monson~\etal{}~\cite{monson2018usability} evaluated two secure email prototypes with users on how deletion can be achieved in secure email communications.

Numerous studies have pointed out several obstacles that make it difficult to delete data completely. Complete deletion methods can be expensive and may render storage unusable. For instance, securely overwriting data on magnetic media may require performing 35 specific-pattern overwrites on each block~\cite{gutmann1996secure}. Scrubbing methods, which involve draining the electrical charge from cells, as explored by Wei~\etal{}~\cite{wei2011reliably}, render the cells unusable after deletion. Moreover, other technologies present their own challenges. For instance, Solid State Disks (SSDs) implement a wear-leveling technique that can prevent the complete overwriting of all cells, potentially leaving deleted data vulnerable to recovery~\cite{singh2016secure}. Machine learning models can memorize the data they have been trained on, but making them forget it is challenging~\cite{song2017machine,veale2018algorithms}. Almuhimedi~\etal{}~\cite{almuhimedi2013tweets} also revealed that Twitter posts are challenging to delete due to replies, comments, and the presence of internet archives. Similarly, Zhou~\etal{}~\cite{minaei2022empirical} showed that users' historical deletion patterns could reveal regrettable tweets. Reardon~\etal{}~\cite{reardon2014secure} also highlighted that secure deletion is challenging because of many different adversarial capabilities.

In cloud deletion, Ramokapane~\etal{}~\cite{ramokapane2016assured} found that despite the intentions of service providers, the inherent characteristics of cloud infrastructure pose significant challenges to achieving secure data deletion. Shu~\etal{}~\cite{shu2017data} also demonstrated that data erasure flaws in the Android system are inherited from the underlying Linux kernel. Encryption-based solutions~\cite{tang2010fade,tang2012secure,ramokapane2016assured} for cloud storage often introduce computational overhead and raise key management concerns. 

\subsection{Users' Perceptions and Practices}
\paragraph{Understanding of deletion}
Regarding users' perceptions and practices of data deletion, numerous studies have revealed that users generally lack understanding or awareness that nominal deletion does not entirely delete data. For example, Murillo~\cite{Murillo2018Press} found that users' understanding of deletion was limited to the interface; they believed that data was deleted if it was not visible to them. Moreover, they found that some users believed data was retained after deletion for business purposes rather than due to technical constraints. Gutmann and Mark Warner~\cite{gutmann2019fight} argued that users often conflate the terms `deleted' and `erased.' Furthermore, after analyzing the operating systems macOS 10.14 and Windows 10, they discovered unclear or incomprehensible information regarding delete and erase functions. This lack of clarity puts data subjects at risk of accidental data breaches when decommissioning storage devices. In a study by Liu~\etal{}~\cite{liu2022your}, participants displayed varied understandings of account deletion, such as revoking authorization. 

In the context of instant messaging apps, Schnitzler~\etal{}~\cite{schnitzler2020exploring} argued that the term ``deleting messages'' was ambiguous for users, leading them to estimate the consequences of their deletion actions inaccurately. This was also alluded to by Abu-Salma et al.~\cite{abu2017obstacles}, who attributed this misconception to misleading feedback from devices. They pointed out that the warning messages users receive from their devices (e.g., iPhone and Nexus) do not specify whether ``all'' the data being deleted refers to application-related data stored on the phone or the data associated with the account on the provider's servers. Also, Liu~\etal{}~\cite{liu2022your} found that users leave their mobile app accounts undeleted because they are unaware of their existence after deleting the apps or that they even need to delete accounts. Regarding cloud deletion, Ramokapane~\etal{}~\cite{Ramokapane2017Feel} found that some users faced difficulties in deletion due to incomplete mental models of the deletion process or how the cloud works. Also, Wermke~\etal{}~\cite{Wermke20Cloudy} found that cloud users are always unsure of the number of copies in the cloud or the procedures to delete such copies.

\paragraph{Reasons for Deletion}
In terms of reasons for deletion, prior studies~\cite{Ramokapane2017Feel,Murillo2018Press,khan2018forgotten} have indicated that users employ deletion as a means to safeguard their privacy. Some users delete data to free up storage~\cite{Ramokapane2017Feel,alhelali2023multiuser}, rectify mistakes~\cite{johnson2012facebook,schnitzler2020exploring}, or eliminate unwanted or outdated information~\cite{Ramokapane2017Feel,khan2018forgotten}. Others~\cite{wang2011regretted,zhou2016tweet,minaei2022empirical} delete as a coping strategy for regrettable content. 

\paragraph{Consequences of Deletion}
Deletion actions can give rise to conflicts and distress. Yilmaz~\etal{}~\cite{yilmaz2021perceptions} discovered that when a social media post evoked positive memories, other users deemed it unacceptable for the owner to delete it. Minaei~\etal{}~\cite{minaei2019conceal,minaei2021deceptive} found that third-party archival services could deduce damaging content from deleted tweets. Regarding deleted posts about themselves, social media users desired privacy for their deleted posts, particularly from large-scale data collectors~\cite{minaei2022empirical}. In a survey by Alhelali~\etal{}~\cite{alhelali2023multiuser}, 13\% of conflicts were linked to deletion, especially where folder members deleted files without consulting others. Ramokapane~\etal{}~\cite{Ramokapane2017Feel} explained that conflicts around shared folders can also arise from members' lack of awareness that deleting an item from the folder removes it for everyone.

\paragraph{Users' Deletion Challenges}
Several studies have identified usability as a hindering factor in data deletion from services. For instance, Minaei~\etal{}~\cite{minaei2022empirical} reported that users found selective deletion mechanisms on social media ineffective in protecting their sensitive deletions. Schaffner~\etal{}~\cite{schaffner2022understanding} found that many social media account deletion interfaces incorporated dark patterns, such as confusing terminologies that could result in accounts not being fully deleted. Previous works~\cite{Ramokapane2017Feel,liu2022your} have shown that users struggle to delete data due to poorly designed deletion mechanisms. Furthermore, Ramokapane~\etal{}~\cite{ramokapane2022users} argued that current cloud deletion mechanisms fail to accommodate users' diverse deletion preferences and emphasized the lack of standardization in deletion-related information. Liu~\etal{}~\cite{liu2022your} found that most individuals who deleted their accounts only read the introduction to account deletion information in privacy policies. Other studies~\cite{habib2019empirical,habib2020s,schaffner2022understanding,liu2022your,take2022feels} have highlighted users' challenges in locating deletion controls across various technologies. For instance, Schaffner~\etal{}~\cite{schaffner2022understanding} and Liu~\etal{}~\cite{liu2022your} found that while account deletion is a common desire, users may abandon the process if they cannot easily locate the deletion option.

\paragraph{Empowerment of Users} Experts have suggested that users need knowledge in six areas to understand data deletion effectively: backend processes, time duration, backups, derived information, anonymization, and shared copies~\cite{Murillo2018Press}. Schnitzler~\etal{}~\cite{schnitzler2020exploring} reported that users could make more informed decisions about the effects of deletion mechanisms in instant messaging apps if the deletion types available were explained more clearly. Ramokapane~\etal{}~\cite{ramokapane2022users} have advocated for developing improved ways of communicating data deletion to users.

\section{New Paradigm }
\label{Sec:NewParadigm}

\subsection{Explainable Deletion (ExD)}

\begin{figure}[!htp]
	\centering
	\includegraphics[scale=0.55]{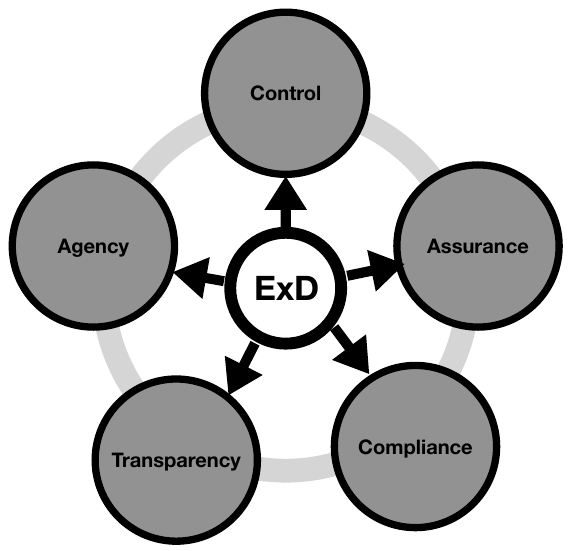}
	\caption{Dimensions of ExD}
	\label{Fig:Dimensions}
\end{figure}

\begin{framed}
 Explainable Deletion is \textbf{\textit{a collection of methods and techniques that provide agency, control, compliance, assurance, and transparency concerning data deletion in a given system.}}
\end{framed}

As per our definition, we argue that the dimensions of~\ac{ExD} are not orthogonal; they are interconnected and dependent on each other. This means that achieving explainable deletion requires finding the right balance or optimal point that considers all of these dimensions simultaneously rather than focusing on them in isolation. For instance, simply providing more technical details may not guarantee user agency, while offering clearer assurance markers with less technical complexity may be more helpful~\cite{molnar2020interpretable}. Drawing parallels to e-voting systems~\cite{Crowe2020Internet}, one of the main challenges they face is the lack of trust markers. In traditional physical voting, numerous trust markers such as booths, sealed boxes, and observers during unsealing boxes and counting contribute to transparency and trust. On the other hand, e-voting systems rely on transparency through open algorithms and mathematical proofs. However, these technical aspects are often not easily understandable by experts and citizens, leading to trust issues.

\emph{Explainable Deletion} represents a novel paradigm that seeks to provide deletion explanations that must address related yet distinct questions concerning data deletion, including aspects such as what, how, when, who, where, and why, tailored to a specific entity. Our vision for deletion explanations is to offer diverse perspectives that cater to specific deletion aspects that are meaningful to the user. It is not necessary for deletion explanations to cover all these aspects, as the explanations themselves may prioritize certain dimensions over others. Moreover, not all users will require or benefit from all explanations, so providing every explanation for every system is unnecessary. We argue that deletion explanations should have a clear objective and be provided to relevant users only when necessary to avoid complicating their interaction with the system. These explanations should be valuable to the user, enabling them to make informed decisions and take appropriate actions based on the information.

For instance, consider the scenario of deleted files resurfacing. A regular user may wonder, ``Why have my files reappeared after I deleted them years ago?'' This experience may create a conflict between their expectations and reality, contradicting their existing knowledge. However, this phenomenon may not surprise an expert user, who may attribute it to the system's deletion method or the level of completeness in the deletion process. In this case, both parties need to know \emph{why} it happened, but providing them with the same explanation may not be ideal. It is essential to balance comprehensibility and usefulness, which may differ for each party.

Next, we will discuss the specific aspects of data deletion that deletion explanations may aim to address.

\paragraph{What - Explanations}
The ``what'' component of deletion explanations focuses on the type of data the system is capable of, intended to, or has already acted upon. Under this component, explanations provide the user with information regarding the data that can potentially be deleted within a system, the data that a deletion request will impact, or the data the system has already deleted. 

\paragraph{How - Explanations}
The ``how'' aspect outlines the deletion procedure, including the specific methods employed by the system to delete data. Explanations pertaining to the ``how'' explain the system's internal mechanisms responsible for fulfilling the deletion request. These explanations provide the user with insights into deletion procedures, such as how all copies of the data and related metadata are deleted or if the deletion is irreversible with no possibility of recovery.

\paragraph{When - Explanations}
These explanations aim to address questions regarding time, providing clarity on when the data will be deleted, the duration of the retention period, and the estimated time required to complete the deletion of data successfully. By offering insights into the timing aspects, the explanations effectively bridge the gap between when the user request for deletion and when the actual deletion procedure takes place or completes.

\paragraph{Who - Explanations}
The ``who'' explanations provide information about the parties involved in the data deletion process and their responsibilities. They clarify who is authorized to request data deletion and who will be responsible for the necessary actions. For instance, an explanation may provide information on individuals or entities with the authority to delete data permanently, or information on the entities which will need execute the deletion procedure for deletion to complete successfully.

\paragraph{Where - Explanations}
The ``where'' explanations are centered around the aspect of the location, providing information about the whereabouts of certain elements. This could include data, controls, or guide. ``Where" explanations may offer insights into where specific deletion controls can be found or indicate where deleted data may be stored or found (i.e., recycle bin).

\paragraph{Why - Explanations}
``Why'' explanations seek to assist users in understanding the behavior of the system prior to, during, and following the deletion request. They shed light on the underlying reasons why the system behaves in a particular way. For instance, a user might get a message explaining why the free cloud storage quota has not changed after deleting some files, or the reason why other users are still able to interact with a deleted social media post. 

\subsection{Explainable Deletion for whom?}
We envision explainable deletion as having relevance and benefit for various individuals and groups involved or having an interest in data deletion processes and implications. This includes not only users but also other entities involved in developing or using technology. In this section, we discuss these different groups. By doing so, we emphasize that explanations for data deletion can vary depending on who is involved and their specific needs. For example, explanations intended for system users may differ in context and length compared to those aimed at regulatory bodies.

\paragraph{Users}
Users of various systems that collect and delete data are critical stakeholders in explainable deletion. We see system users as having a direct interest in understanding data deletion and the mechanisms to delete data from systems.

\vspace{5pt}
\begin{mdframed}[backgroundcolor=gray!10,linewidth=0.1pt,font=\small\sffamily] 
        \textbf{Example.} There has been an increase in the number of paid and free services (e.g., DeleteMe~\footnote{https://joindeleteme.com/}, Kanary~\footnote{https://www.kanary.com/how-it-works}), that offer guidance and assistance in removing personal data online~\cite{take2022feels}. Moreover, a high number of users are seeking information on how to exercise their right to delete their data. For example, from 2015 to 2021, more than 255 000 people in France exercised their right to delete the ``right to be forgotten'' to Google and Bing, according to SurfShark~\footnote{https://www.statista.com/statistics/1373747/right-to-be-forgotten-total-requests-europe-by-country/}. These trends highlight the growing interest among users in understanding and achieving data deletion.
\end{mdframed}
   
\paragraph{Service Providers}
Entities and organizations that develop technology or provide services to users should also be interested in explainable deletion. Service providers are responsible for complying with regulations, particularly those related to data deletion~\cite{CCPA2023Consumer,GDPR2023Deletion}. Moreover, they are also responsible for providing deletion mechanisms in their system to allow users to delete their data. Lastly, providing a transparent system would benefit them business-wise, fostering trust and confidence in their customers and partners~\cite{ras2022explainable}. 

\vspace{5pt}
\begin{mdframed}[backgroundcolor=gray!10,linewidth=0.1pt,font=\small\sffamily] 
    \textbf{Example.} According to Statista~\cite{Statista2019Consumer}, it has been reported that one of the most challenging obligations for companies in the U.S. and Europe to comply with in 2019 was the GDPR right to be forgotten. More recently, there have been debates about how ChatGPT handles inaccurate or misleading information and inquiries regarding its compliance with the GDPR right to be forgotten. Reports (e.g.,~\cite{Uri2023ChatGPT}) continue to say that currently, ChatGPT does not provide procedures for individuals to request data deletion.
\end{mdframed}

\paragraph{Regulators}
We also anticipate regulatory bodies responsible for ensuring that organizations and companies offering services comply with data protection and privacy regulations being interested in explainable deletion. These entities would want to ensure that organizations and companies adhere to the prescribed guidelines concerning data deletion practices.

\vspace{5pt}
\begin{mdframed}[backgroundcolor=gray!10,linewidth=0.1pt,font=\small\sffamily] 
\textbf{Example.} Many countries are now following the E.U. (i.e., GDPR) and California (i.e., CCPA) by establishing laws focusing on data deletion. As regulators enact these laws, they will be interested in how companies comply with the requirements or the mechanisms they employ to show compliance. Australia is the latest to consider the right to delete in their privacy act~\cite{Pual2023Guardian}.
\end{mdframed}

\paragraph{Developers}
Designers and application developers who build systems should also be interested in explainable deletion as they need to build compliant systems. They need to know how to implement deletion explanations or what to do to meet the regulatory requirements.
\vspace{5pt}
\begin{mdframed}[backgroundcolor=gray!10,linewidth=0.1pt,font=\small\sffamily] 
\textbf{Example.} Considering the need to build compliant applications, developers will be interested in~\ac{ExD}. Previous research~\cite{abdalkareem2017developers,meng2018application,tahaei2022charting} has argued that while most systems are not compliant with regulations and many developers lacking understanding of compliance, most developers are interested in complying with regulations. \ac{ExD} would provide developers with methods and techniques to achieve compliance. In his article for REUTERS, Bellamy~\cite{Fredric2023Laws} argues that the new era requires an understanding of laws and that this understanding will create a foundation from which to analyze and comprehend requirements.
\end{mdframed}

\paragraph{Researchers}
We also see academics and industry researchers having an interest in explainable deletion. They may be interested in exploring and developing novel approaches, frameworks, and best practices for implementing deletion explanations. 
\vspace{5pt}
\begin{mdframed}[backgroundcolor=gray!10,linewidth=0.1pt,font=\small\sffamily] 
\textbf{Example.}There is a significant amount of research focusing on explainable AI; however, there is a need for empirical evidence, particularly on ~\ac{ExD}. In Section~\ref{Sec:Agenda} of this paper, we discuss the gaps that should spark interest from researchers.
\end{mdframed}

\paragraph{Legal Experts}
Deletion explanations can be useful to legal professionals specializing in data protection and privacy laws. They can use explanations to ensure clients adhere to regulations and respect users' rights. 
\vspace{5pt}
\begin{mdframed}[backgroundcolor=gray!10,linewidth=0.1pt,font=\small\sffamily] 
\textbf{Example.} In 2014, there was a legal case~\footnote{https://eur-lex.europa.eu/legal-content/EN/TXT/HTML/?uri=CELEX:62012CJ0131} between Google Spain and Google Inc. versus the Spanish Data Protection Agency (AEPD) regarding Mario Costeja Gonzalez's search results. The AEPD requested that Google remove inaccurate data related to Mario from the Google search results. Commentators have pointed out that while search results may be denied in the EU, it does not necessarily mean the search results are completely deleted. This case highlights the importance of understanding the legal aspects and how they translate into technical implementation.
\end{mdframed}

\paragraph{Privacy Advocates}
Lastly, deletion explanations can be valuable to individuals and organizations advocating data privacy rights. These advocates emphasize the importance of transparency and accountability, and deletion explanations may serve to achieve these objectives. 
\vspace{5pt}
\begin{mdframed}[backgroundcolor=gray!10,linewidth=0.1pt,font=\small\sffamily] 
\textbf{Example.}  Recently, many advocacy groups have supported various groups and individuals in understanding how to delete their data. For instance, in 2023, privacyrights.org~\footnote{https://privacyrights.org/resources/california-delete-act-bill-give-californians-more-control-over-their-personal-data} sponsored the California Delete Act (S.B)~\cite{DeleteAct2023Consumer}, which aims to provide residents of California with essential tools to take control of their personal information and protect their privacy. Other organizations, such as PrivacyDuck~\footnote{https://privacyduck.square.site/}, would also benefit from~\ac{ExD} since they offer free and paid resources to users. Media outlets like BusinessInsider~\cite{BusinessInsider2023Consumer} have also published guides on deleting data online, further promoting awareness and education about deletion. They need to know how to delete to help others delete.
\end{mdframed}

\section{How to Explain Deletion}
\label{Sec:HowTo}

We posit that the foundation of designing deletion explanations lies in providing high-quality information about data deletion in a meaningful and useful manner to the intended audience. A useful deletion explanation is one that meets a certain standard of what constitutes a satisfactory explanation based on the recipient, usage conditions, and the specific task at hand. Generally, the criteria for a satisfactory explanation are typically qualitative and not quantifiable. Previous research~\cite{friedrich2011taxonomy,hosseini2016foundations,carvalho2019machine,chazette2020explainability,liao2020questioning,chazette2021exploring} has proposed several desirable properties of satisfactory explanations. For instance, Chazette~\etal{}~\cite{chazette2020explainability} emphasized that explanations should fulfill criteria such as accessibility, usability, informativeness, understandability, and auditability to achieve system transparency. Other studies~\cite{friedrich2011taxonomy,hosseini2018four,ras2022explainable} suggest that the quality of explanations is dependent on the user and the specific context. This implies that deletion explanations should not aim to be one-size-fits-all; rather, they should be tailored to the individual user and the technology they interact with. In terms of deployment, the implementation of explainable deletion should be tailored to accommodate the distinct characteristics of various systems. Variations in interfaces and technical constraints might necessitate the utilization of diverse methods for delivering explanations. Nevertheless, it is crucial to ensure that the core topics and concepts covered in the explanations remain consistent across all systems. This section discusses various design aspects that should be considered when developing deletion explanations. We recognize that efforts must be broadly focused on three design aspects: content, presentation, and usability.

\subsection{Content} This category pertains to the information that should be included in deletion explanations. Deletion explanations should primarily address the process of data deletion or how the system handles the deletion of data. They should be communicated using language that is accessible and understandable to users, enabling them to make informed decisions regarding data deletion and privacy. The content should be informative, reducing ambiguity and clarifying any uncertainties about the actual deletion process or user expectations. Users normally focus on high-level functionality rather than intricate technical details, so the explanations should be tailored to the user and the specific context. Consistency in content is crucial to establish reliability and trust. For example, an explanation regarding data retention for a particular service should remain consistent regardless of how users access it. Designers should also ensure that the explanations are user- and context-specific to provide only the necessary information for understanding.

\subsection{Presentation} This category focuses on how deletion explanations should be delivered to the receiver, including the delivery method and modality. To promote inclusivity, deletion explanations should be made as accessible as possible. Designers should prioritize making users aware of the existence of these explanations and ensuring their accessibility. Explanations can be presented to the receiver in different modalities, such as textual, audio, or graphical formats. For example, this could be facilitated through existing privacy label frameworks~\cite{kelley2009nutrition,shen2019iot,emami2020ask,koch2022keeping}. While mobile apps and IoT device labeling frameworks are still in their infancy stages, they can be used to present deletion explanations. Various symbols can be designed to represent various deletion concepts and types. Integrating data deletion indicators into existing label frameworks will allow for a more holistic representation of data management practices. The choice of modality should be based on the user and the specific context, focusing on what will effectively explain deletion concepts to the receiver. For example, using privacy labels to represent deletion explanations for IoT devices might work better for users.

Deletion explanations can be delivered to users automatically or upon request. Automatically delivered explanations can be integrated into the user's interaction with the technology and presented when needed. It is essential to ensure deletion explanations do not disrupt user journeys or interfere with their primary tasks. The other option is to allow users to request or invoke an explanation. Explanations should be placed where users can easily find them when needed. In the earlier example of Dropbox, users could have been allowed to visit a specific deletion page to find additional information about how Dropbox handles data deletion. Identifying the appropriate locations for these explanations is crucial, and designers should choose places where users are most likely to search for such information. Currently, users struggle to find information about data deletion~\cite{Ramokapane2017Feel,Murillo2018Press,habib2019empirical,ramokapane2022users}. Signposting them to these locations is also essential. Allowing users to request explanations when needed can prevent them from hindering their primary tasks. The mechanisms used to relay the information should give the user control, allowing them to skip an explanation if desired.

When determining how much information should be provided, explanations should be concise and succinct to prevent cognitive overload and overwhelming the user. Reusing explanations, when applicable, can help reinforce understanding and beliefs, as well as maintain consistency. Moreover, by reusing explanations, fidelity is preserved, and users can develop a consistent mental model of deletion processes.

\subsection{Usability} This category emphasizes the importance of considering the usefulness and comprehensibility of deletion explanations from the users' perspective. The goal of an explanation is to facilitate users' understanding of data deletion, which requires tailoring the content to their specific needs and level of expertise. Therefore, transparency in deletion will become meaningful if it enhances users' comprehension, ensuring that the information provided is understandable and accessible to the intended audience.

Furthermore, the usefulness of the information lies in its potential to influence users' perceptions and guide their actions. Deletion explanations should empower information receivers to make informed decisions and act appropriately. To achieve this, explanations should address relevant questions and omit irrelevant or redundant information that users may already be familiar with. For instance, an explanation designed to guide users on locating deletion controls should not go into detailed instructions on how to use those controls. We are not advising against multiple explanations being presented together. We are only suggesting that when multiple explanations co-exist, they should be logically connected and presented to the user for a specific purpose.

Consideration should also be given to users' existing background knowledge and beliefs. Building upon their existing understanding of deletion processes can enhance acceptance and prevent misconceptions. Previous research~\cite{venkatesh2003user} suggests that introducing new information that contradicts or does not align with users' beliefs can create resistance. Therefore, explanations should be designed to align with users' mental models and facilitate the development of proficiency in deleting data from various systems.

Usability also entails providing users with control and interactivity when possible and appropriate. Users should be able to engage with the explanation according to their needs, such as requesting additional information for clarification. By allowing users to interact with the explanation, it becomes more personalized and responsive to their specific requirements.

Overall, deletion explanations should be designed using a user-centric approach, considering users' needs and mental gaps, as their acceptance and usefulness depend not only on their accuracy (or correctness) but also on other various factors such as users' background, the context of use, experience, and the specific technology they seek to understand. Recognizing the diverse range of users and their unique requirements is crucial in creating explanations that effectively convey information and foster understanding. By tailoring explanations to align with users' perspectives and providing information that resonates with their cognitive abilities and prior knowledge, the overall quality and effectiveness of the explanations can be significantly enhanced.

\begin{figure*}[!htp]
	\centering
	\includegraphics[scale=0.45]{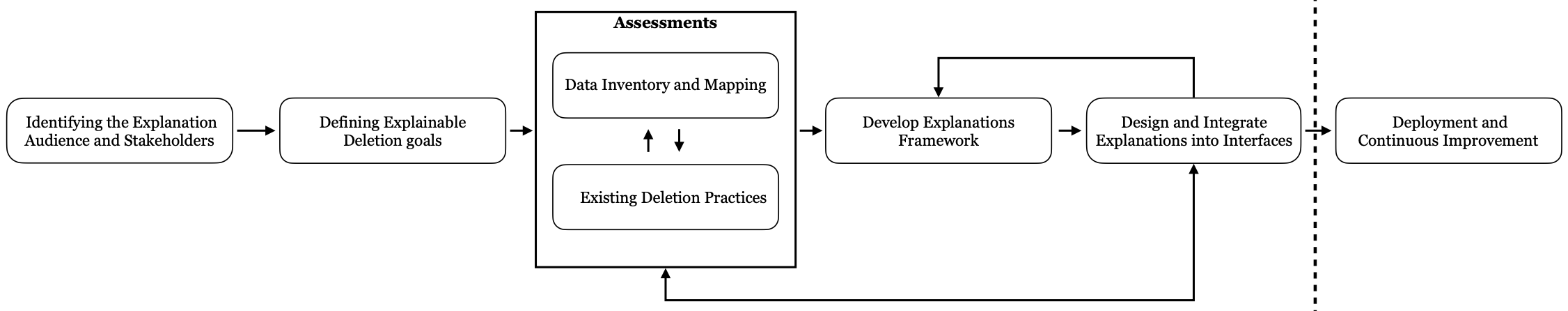}
	\caption{Summary of various steps for operationalizing \ac{ExD}.}
	\label{Fig:Framework}
\end{figure*}

\section{Operationalizing Explainable Deletion}
\label{Sec:Operationalising}

We envision~\ac{ExD} to be an integral component of the system development cycle. Designers and developers should consider the concept of explainable deletion during the initial stages of system conceptualization and creation. This approach not only allows for the integration of user-centric design principles but also promotes effective data management and proactive compliance efforts. For example, designers can leverage this opportunity to create interfaces and processes that align with users' needs and preferences right from the outset. While we believe that the optimal time to operationalize ~\ac{ExD} is during the design and developmental phases emphasizing privacy-, explainable-, and transparency-by-design, it can also be effectively deployed in existing systems. We do not advocate for dismantling existing systems; rather, we emphasize that~\ac{ExD} can be seamlessly integrated into established systems. Integrating~\ac{ExD} into existing systems ensures ongoing good data management, compliance with regulations, and the rebuilding of user trust. Furthermore, deploying~\ac{ExD} is not a one-off effort but rather an iterative process that requires revisiting and improvement as requirements and regulations around data deletion evolve.

Figure~\ref{Fig:Framework} depicts the different steps involved in operationalizing~\ac{ExD}. The following sections provide a comprehensive description of these steps:

\paragraph{(1) Identifying the explanation audience and stakeholders.}The initial step entails identifying the target audience or the specific group for whom the explanations are being crafted. The selection process is crucial because explanations should be personalized and not too generalized. However, this is not to say that explanations cannot serve more than one group or be generalized. The second category to identify comprises the experts who will actively engage in designing and implementing deletion explanations. It is important to note that this process is not exclusive to system designers; it encompasses anyone with data deletion expertise, such as legal teams, developers, and user experience (UX) experts.

\paragraph{(2) Defining Explainable deletion goals.}
The second step revolves around outlining the objectives of the deletion explanations. These goals should outline what the explanations aim to achieve by making data deletion processes more transparent and comprehensible. It is vital that these goals are aligned with the objectives of the audience engaging with the system or service.

\paragraph{(3) Data inventory and Mapping.} Progressing to the subsequent stage involves carefully identifying and compiling various data types collected, shared, or generated by the system. This inventory should also encompass the data storage locations or potential points of passage. During this stage, it is also vital to identify the corresponding deletion controls and mechanisms employed for deleting such data.

\paragraph{(4) Assess Current Deletion Practices}
For deployed systems, this stage centers on evaluating the existing deletion practices and associated informational elements. It begins with assessing how the data identified in the preceding stage is currently being deleted. It also involves scrutinizing the alignment between deletion controls, policies, and the data that can be deleted. Furthermore, it aims to determine if the provided deletion information sufficiently explains data deletion and associated controls, thereby identifying any informational gaps. In cases where the system is yet to be launched, this phase may extend to ensuring all data within the system is accounted for in terms of deletion controls, mechanisms, and information.

\paragraph{(5) Develop Explanations Framework.}
Guided by the data inventory and data deletion assessment outcomes, the next step involves creating an encompassing explanation framework that addresses various dimensions of data deletion. This should be a collaborative effort that includes all stakeholders (i.e., experts) and aims to clarify data deletion of all collected and shared data. Explanations should attempt to explain the `what,' `how,' `when,' `who,' `where,' and `why' aspects of deletion. During this stage, prioritizing the comprehensibility and usability of content (i.e., explanations) is essential, and this should be based on the intended target audience. As explanations are developed, they should undergo testing with the intended audience to ascertain their clarity. Introducing new features post-explanation deployment should trigger the development of corresponding new explanations.

\begin{figure*}[!htp]
	\centering
	\includegraphics[scale=0.75]{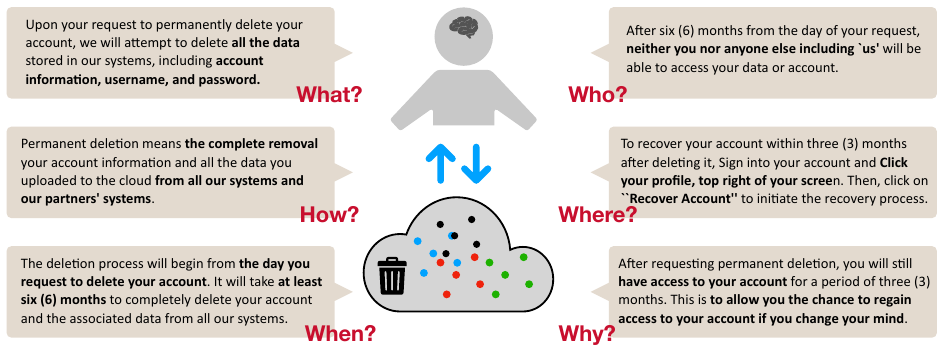}
	\caption{This is sample of how deletion explanations may look like for the deletion of a cloud account. With the application of \ac{ExD} the user has a better chance of understanding what will happen when they request for their account to be deleted.}
	\label{Fig:Example}
\end{figure*}

\paragraph{(6) Design and Integrate Explanations into Interfaces.}
Following the formulation of the framework, the following stage involves crafting user-friendly interfaces (e.g., voice, graphical) that meet audience' expectations, concerns, and preferences concerning data deletion explanations. During this phase, decisions regarding the presentation and usability of the interfaces should be made. The designs should seamlessly integrate with the existing user experience while causing minimal disruption to established system functionality.

\paragraph{(7) Testing and Evaluating Explanations with the Relevant Audience.}
This phase is intertwined with the previous two steps, the design, development, and testing of the explanations and interfaces conducted with the intended audience. Should testing not yield satisfactory outcomes (e.g., explanations being hard to understand or the interfaces affecting the system's usability), revisiting the design and development of explanations should become necessary.

\paragraph{(8) Deployment and Continuous Improvement.}
Upon successfully meeting testing requirements, the explanations can then be deployed. Nonetheless, our viewpoint emphasizes that developing and deploying data deletion explanations should not be a singular exercise. Teams must continuously refine and enhance deletion explanations, informed by user feedback, integration of new features, evolving best practices, and shifts in data protection regulations.

\subsection*{Example: Permanently Deleting a Cloud Account}
A simplified example of an explainable deletion framework for permanently deleting a cloud account is presented in Figure~\ref{Fig:Example}. In this illustration, the intended audience is the users of the cloud platform to provide them agency, control, and assurance when engaging with the deletion process offered by the cloud service provider. We assume the user intends to delete their account entirely, including all associated data. It is also assumed that the user has successfully located the deletion controls to initiate the process of permanent account erasure. The exemplar provides high-level information in textual form, designed to help users in comprehending the procedure and outcomes of deleting their cloud account. In practice, this content would be identified through a series of empirical inquiries involving relevant stakeholders such as system designers, legal teams, and users.

\section{Research Agenda}
\label{Sec:Agenda}
While providing explanations for deletion can benefit users, it may also introduce new challenges. In this section, we explore the potential challenges that can arise due to explaining deletion. Based on these challenges, we draw up a research agenda for future studies aimed at promoting deletion transparency in systems.

\subsection{Adversarial Tactics}
Adversarial tactics pose challenges to the realization of explainable deletion by exploiting vulnerabilities in the system and undermining the transparency and effectiveness of the deletion process. Certain techniques may hinder deletion, making deletion explanations untrustworthy and unreliable. For example, adversaries may attempt to evade deletion by exploiting system vulnerabilities or manipulating the logic (using dark patterns to discourage deletion~\cite{schaffner2022understanding}).
 
\paragraph{Research Agenda.} Researchers should focus on developing new threat models that consider various adversaries and their motivations regarding deletion. This will help identify potential vulnerabilities, develop robust security measures to protect systems, and ensure the credibility of deletion explanations. Future research could also explore methods for detecting and preventing deletion evasion.

\subsection{Commercial constraints or proprietary technologies}
Explaining the internal workings of a system, even for deletion purposes, may not be easy to achieve due to commercial constraints and proprietary reasons~\cite{ras2022explainable}. Service providers may risk losing their competitive edge against other companies. Moreover, \ac{ExD} might be challenging to deliver, given that data often cross many system boundaries.

\paragraph{Research Agenda.} 
Future research is needed to investigate methods that can provide explainability in deletion processes without compromising the business competition of the provider. Research should identify which information is suitable to share and determine the appropriate depth of that information.

\subsection{New Concerns}
Providing detailed explanations about data deletion, for instance, information about what data is or is not deleted, can raise new privacy concerns. The introduction of new information might inadvertently reinforce misunderstandings and biases about systems, which could discourage users from wanting to interact with these systems~\cite{dzindolet2003role, abdul2018trends, dodge2019explaining}. Rather than fostering trust, misunderstanding of certain disclosures may raise doubts or suspicions among users, undermining their confidence in the system's integrity and reliability. 

\paragraph{Research Agenda.} 
Further research is required to examine the influence of deletion explanations on users' trust in systems and to develop strategies for establishing and sustaining trust. Studies can assess trust levels before and after providing explanations, providing valuable insights into the impact of transparency. This calls for more research concerning misconceptions around data deletion, prior work~\cite{Ramokapane2017Feel} found that users fail to delete in the cloud because of their incomplete mental models. There is a need to explore users' expectations, concerns, and requirements concerning deletion processes in various technologies (e.g., smart home devices). A comprehensive understanding of users' expectations, concerns, and needs will inform the design and implementation of effective deletion mechanisms across various technologies. While there are existing efforts aimed at understanding the deletion of various technologies, such as cloud~\cite{Ramokapane2017Feel,khan2018forgotten,ramokapane2022users} and web technologies~\cite{Murillo2018Press,liu2022your}, a knowledge gap exists concerning users' comprehension of deleted data, associated risks, and preferences. Future studies should investigate users' understanding of deleted data, the risks involved, and their preferences regarding it.

\subsection{Cost implications}
Explaining deletion can be resource-intensive and costly for system owners. Designing and implementing comprehensive explanations may require significant investments in terms of time, effort, and financial resources~\cite{wang2019designing}. Moreover, there may be a need to acquire new skills or expertise to effectively communicate the complexities of data deletion to users. These factors contribute to the potential cost implications of implementing explainable deletion practices.

\paragraph{Research Agenda.} 
    Future research should assess the skills and expertise needed to provide deletion explanations effectively. There is also a need for interdisciplinary research to identify and promote best practices for cost-efficient and effective implementation of explainable deletion. There could be some collaborative effort with various stakeholders to develop guidelines, standards, or frameworks to help organizations navigate the cost implications while ensuring transparency and accountability in data deletion practices. Lastly, system owners should evaluate the benefits of gaining trust and regulatory compliance against the cost of implementing the \ac{ExD}.

\subsection{User overload}
Introducing deletion explanations within modern systems may aggravate the existing complexities and overload users with additional information, potentially worsening their overall user experience. Offering detailed explanations might overwhelm users, affecting their ability to navigate and comprehend the system effectively~\cite{coppers2018intellingo,ras2022explainable}.

\paragraph{Research Agenda.} 
    Future research could focus on strategies to mitigate information overload and ensure effective user understanding. This can include exploring techniques such as friendly visualizations or summaries to reduce cognitive load and enhance comprehension. Adopting a user-centered design approach for deletion explanations can ensure intuitive interfaces that meet user needs and are part of user experience~\cite{zhou2016making}. Additionally, there is a need to determine the appropriate context and timing for delivering explanations to maximize their usefulness to users.


\subsection{Challenges of explaining technical concepts} 
Explaining technical operations accurately, comprehensively, and briefly in a simple and non-technical way presents a significant challenge~\cite{coppers2018intellingo,ras2018explanation,ras2022explainable}. Deletion processes, in particular, can be inherently complex and may need specialized technical knowledge to comprehend and effectively explain to others fully. Additionally, the complex nature of certain infrastructures, including their data storage and deletion mechanisms, further complicates explaining their behavior to designers and end users.

\paragraph{Research Agenda.} 
    Future work could aim to understand the complexities of different data storage and deletion mechanisms to identify key challenges and develop easy-to-understand strategies for explaining their behavior. Research could analyze existing infrastructures, protocols, and standards to uncover areas where explanation gaps exist and propose solutions. There is also a need to explore methods that can simplify technical operations and complex processes, making them comprehensible to relevant parties.

\subsection{Criteria for Good Explanation}
Defining the criteria for a good explanation might pose another serious challenge~\cite{hoffman2017explaining,klein2018explaining}. Should explanations prioritize providing the correct answer or ensuring that users can understand the answer? The former often disregards the listener's level of understanding and assumes there is only one correct explanation. However, the latter is more pragmatic, considers the audience, and aims to provide explanations that are easy to understand~\cite{kim2018explainable}. While we argue for pragmatic deletion explanations, it is essential to acknowledge that such explanations may introduce confusion and potentially include incorrect information. Resolving these issues presents a significant challenge in achieving explainable deletion.

\paragraph{Research Agenda.} 
    There is a need for user-centric explanations; research should engage users to understand their preferences and needs regarding deletion explanations. Ramokapane~\cite{ramokapane2022users} investigated users' needs concerning deletion information for the cloud. Future works should extend that work and investigate how various groups perceive and evaluate explanations. There is also a need to balance correctness and understandability. Explanations should contain information that is correct and understandable by users.  

\subsection{Implementation Challenges}
Lastly, deletion explanations might be technically challenging to implement. Incorporating these explanations into existing systems can be technically demanding. Designing new interfaces and user interactions to accommodate deletion explanations may pose challenges for developers. Moreover, integrating explanations seamlessly into user journeys may result in increased complexity.

\paragraph{Research Agenda.} 
    Future research should investigate how deletion explanation could be integrated into existing systems and user journeys without disrupting the system functionality and user experience. Future works also need to address scalability and performance issues. Lastly, developers need to be supported; tools, libraries, APIs, documentation, and tutorials are all crucial for realizing~\ac{ExD}.

\section{Conclusion}
\label{Sec:Conclusion}
Data has a lifecycle. It is created, stored, used/reused, shared, combined, analysed and sysnthesised. And, finally, it is deleted. Or it should be. After all, this right to removal of one's data from digital systems is enshrined in many legal regulations such as the GDPR. Yet deletion remains shrouded in mystery -- from complexity and usability issues regarding initiating data deletion to understanding what happens as a consequence through to gaining assurance that it has indeed happened and transparency about the processes utilised by the requisite party (or parties) in deleting the data. This leads to inherent and fundamental asymmetry between providers and users. Provision of data is easy yet removal is difficult nigh impossible. In this paper, we have highlighted these very challenges and issues and argued that explainability of data deletion is as critical as clarity over data collection, use and sharing.

Explainable deletion is, however, not without its challenges. Many key facets need to be addressed to deliver this new paradigm -- from design and properties of the explanations to suitable means of presenting and communicating them to different stakeholder groups through to technical measures that deliver transparency, control and assurance. The research agenda we propose highlights a number of directions to develop the empirical basis, technical underpinnings, socio-economic insights and user-centred designs for explainable deletions. These research advances are critical in redressing the aforementioned asymmetry between service providers and users. Without explainable deletion, users will never have true agency over their data. 

\begin{acks}
The authors would like to express their gratitude to Dr. Partha Das Chowdhury for providing thorough feedback on the initial draft of this paper. The work is part of our efforts in making privacy equitable, it has been supported by EPSRC EP/W025361/1. 
\end{acks}

\bibliographystyle{ACM-Reference-Format}
\bibliography{References}

\appendix

\end{document}